\pdfoutput=1
\documentclass{JINST}
\usepackage{textcomp}

\usepackage{amsmath, amsthm, amssymb}
\usepackage[ansinew]{inputenc}
\usepackage{wrapfig}
\usepackage{graphicx}
\usepackage{cite}
\usepackage{url}
\usepackage{float}

\usepackage{xspace}

\mathchardef\mhyphen="2D
\title{The CosmicWatch Desktop Muon Detector: a self-contained, pocket sized particle detector.}
\author{S. N. Axani$^a$\thanks{Corresponding author: saxani@mit.edu}, K. Frankiewicz$^b$, J. M. Conrad$^a$\\
\llap{$^a$}Department of Physics, Massachusetts Institute of Technology\\
77 Massachusetts Av., Cambridge, MA 02139, USA\\
\llap{$^b$}National Centre for Nuclear Research\\
Ho\.za 69, 00-681 Warsaw, Poland\\
\\
E-mail: \email{saxani@mit.edu}}
\abstract{The CosmicWatch Desktop Muon Detector is a self-contained, hand-held cosmic ray muon detector that is valuable for astro/particle physics research applications and outreach. The material cost of each detector is under 100\$ and it takes a novice student approximately four hours to build their first detector. The detectors are powered via a USB connection and the data can either be recorded directly to a computer or to a microSD card.  Arduino- and Python-based software is provided to operate the detector and an online application to plot the data in real-time. In this paper, we describe the various design features, evaluate the performance, and illustrate the detectors capabilities by providing several example measurements.}

\begin{document}
\clearpage
\begin{figure}[h]
\begin{center}
\includegraphics[width=1.\columnwidth]{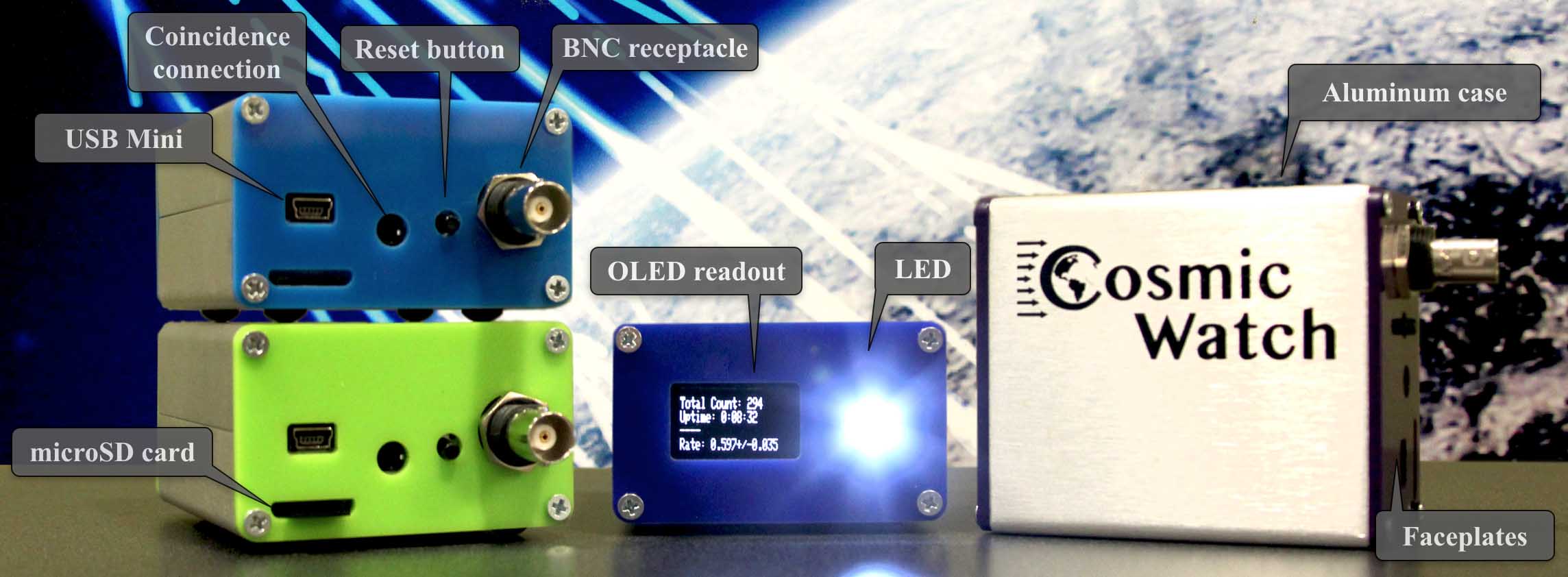}
\caption{An array of the Desktop Muon Detectors.}
  \label{fig:array}
 \end{center}
\end{figure}
\section{Introduction}
This paper describes a very compact muon detector with applications for astrophysics and particle physics, as well as for outreach.  The detector was originally developed  as a prototype sub-detector for the GEN2 upgrade~\cite{GEN2} to the IceCube experiment~\cite{icecube}. It has been applied in various uses in particle physics, including studies of a testbeam at Fermilab that we present in this paper.  It has also received wide recognition as an easy-to-construct educational tool~\cite{ajp}.  Below, we introduce the CosmicWatch outreach program and then discuss this second-generation model of the Desktop Muon Detector, shown in Fig.~\ref{fig:array} .

Similar devices have been used in particle physics experiments, such as the calibration scintillator cubes installed in the MiniBooNE neutrino experiment at Fermilab~\cite{muoncubes} or the RaySure silicon diode microdosimeter (a compact radiation sensor) used in the Rad-X experiment at NASA~\cite{NASA}.
There are also muon detector based outreach projects, like QuarkNet~\cite{QuarkNet}, which aim at providing research experience of high school teachers. Thus, development of the Desktop Muon Detector is a realistic exercise for students who intend to participate in particle physics and astrophysics experiments in the future.

\subsection{The CosmicWatch program}
CosmicWatch is a Massachusetts Institute of Technology (MIT) and National Center for Nuclear Research (NCBJ) based outreach program aimed at introducing students to astro/particle physics while giving them experience in the electronics and machine shop. We have worked with over 50 individuals (including students, teachers, and enthusiasts) to build over 100 cosmic ray muon detectors and have provided resources to many international institutions (US, Poland, Peru, Canada, India, ...). People participating in the program build their own desktop muon detectors and gain experience in working with the detectors, measuring properties associated with cosmic ray muons, programming micro-controllers, developing data analysis techniques, and soldering surface mount components. Further details can be found on the project website, Ref.~\cite{cosmicwatch}. 

\subsection{The Desktop Muon Detector}
The Desktop Muon Detector consists of a 5$\times$5$\times$1~cm$^3$ extruded slab of plastic scintillator instrumented with a silicon photomultiplier (SiPM) -- a light-sensing device. When a charged particle deposits energy in the scintillator, some of that energy is re-emitted isotropically in the form of photons.  Photons incident on the photosensitive area of the SiPM can induce a Geiger discharge in the SiPM micro-cells, creating a measurable current. The produced current is sent through a custom designed printed circuit board (PCB), which amplifies and shapes the signal such that a micro-controller (Arduino Nano~\cite{Arduino}) can measure the event time stamp and peak voltage. If the signal peak voltage is above the software defined threshold, the micro-controller records the event data either to a microSD card or directly to a computer through a USB connection. The CosmicWatch website also provides an online application to plot data in real-time or to record data (see ``Start Measurement" section on the CosmicWatch website~\cite{cosmicwatch}). For each event, the detector records the event number, event time, average measured 10-bit ADC value, calculated SiPM peak voltage, total deadtime, and temperature.

\begin{figure}[t]
\begin{center}
\includegraphics[width=1\columnwidth]{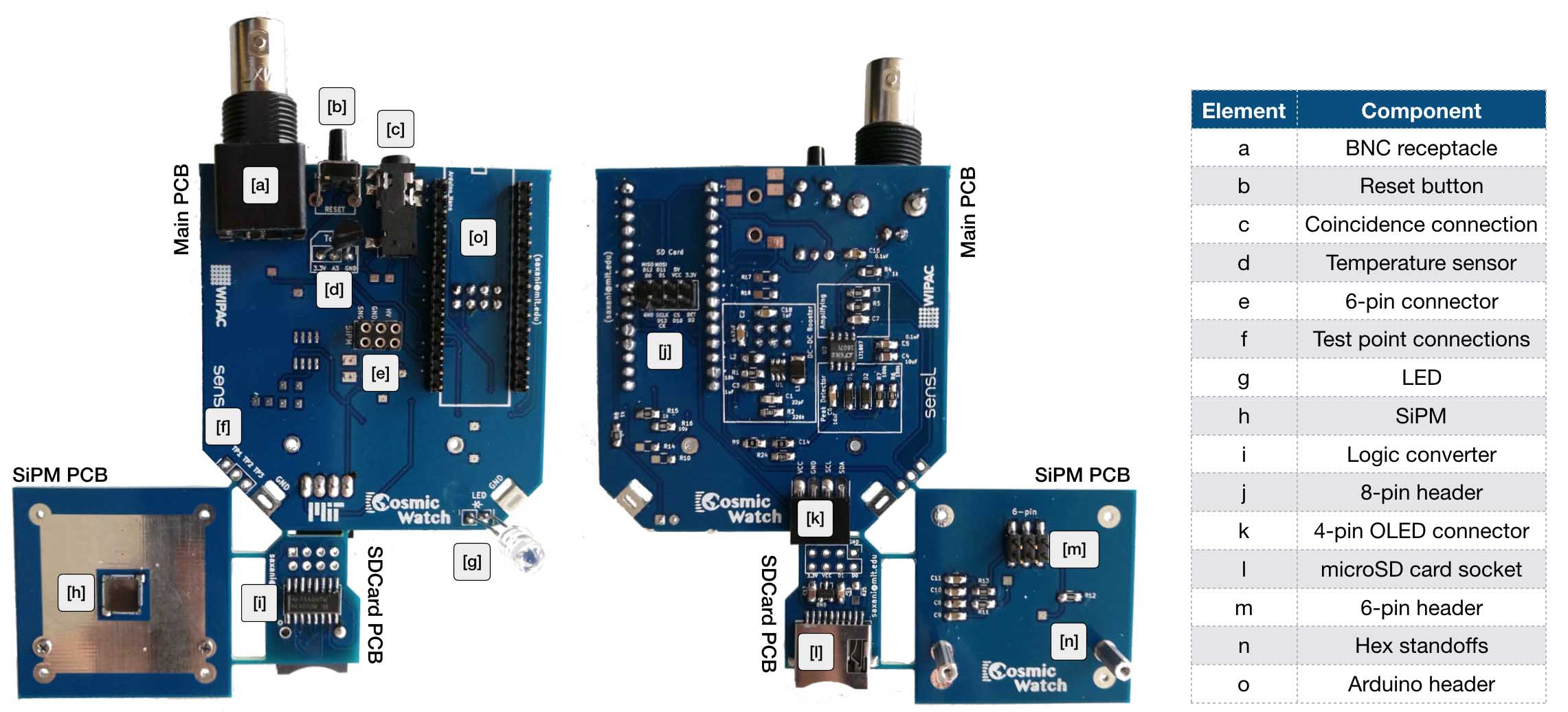}
\caption{The top (left) and bottom (right) of the PCBs. They are two-layer boards and printed together to fit inside a 10$\times$10~cm$^2$ square. All components in this figure have been populated except for the Arduino Nano, which mounts on the rails, element [o].}
  \label{fig:pcb}
 \end{center}
\end{figure}

The detector includes a built-in OLED screen which displays the count number, total uptime, count rate (compensated for the detector deadtime), and an indication bar whose length is proportional to the calculated SiPM peak voltage of the last triggered event.  Multiple detectors can also be linked together via a 3.5~mm male-to-male audio cable to make coincidence measurements. The detector was measured to draw approximately 0.5~W and can be powered either through a USB port on a computer or USB power bank. The total weight, including the aluminum case, is 178~g and the outer dimensions of the detector are 66.4~mm$\times$101.6~mm$\times$39.9~mm (including the protruding BNC receptacle and LED holder). Excluding the aluminum enclosure and endplates, the detector has a weight of approximately 76~g. 

The supplementary material pertaining to this project can be found in the GitHub repository located in Ref.~\cite{github} and will be referred to as needed throughout this paper. Included in the supplementary material is a document (Instructions.pdf) that describes the detailed process of building, testing, and troubleshooting the detector.

\section{Detector design}\label{design}
\hspace{4ex}
The detector is built around three custom designed printed circuit boards: the main PCB, SiPM PCB, and SDcard PCB (shown in Fig.~\ref{fig:pcb}). They connect together via headers and have all been designed to reduce noise by eliminating all wired connections, minimizing track length between components, and strategically placing ground fills to minimize light leaks. The custom PCBs reduces the total cost of the detectors and give students experience mounting 0805 (0.08~inches by 0.05 inches) surface mount components. The shape and size of the PCBs were designed such that they would fit into a small, commercially available aluminum electronics enclosure. The front and back plate for the electronics enclosure are laser cut from 2.5~mm thick acrylic.

\begin{figure}[b]
\begin{center}
\includegraphics[width=1\columnwidth]{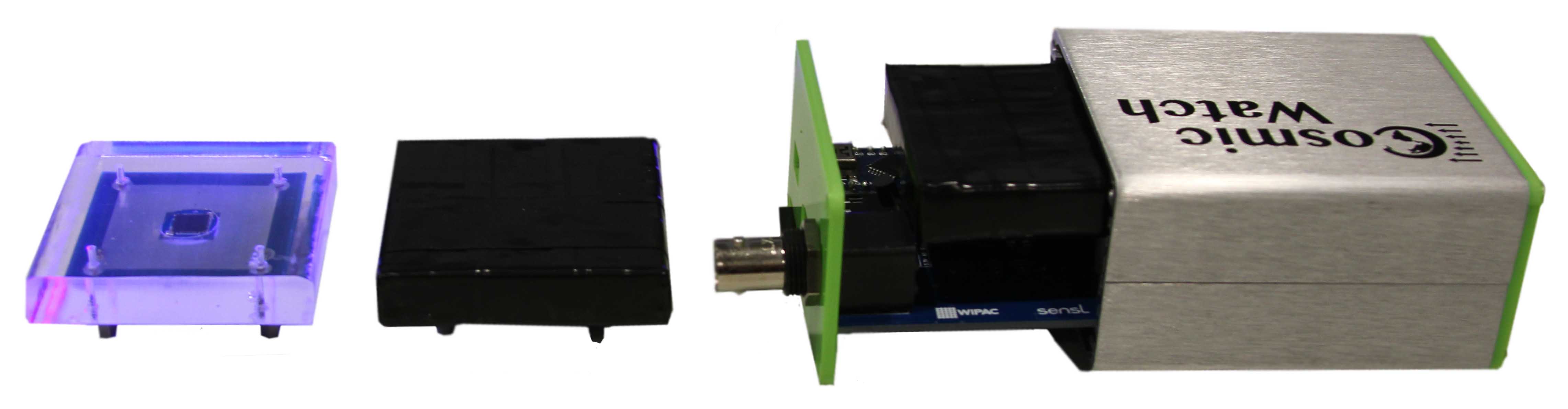}
\caption{Left: the scintillator mounted on the SiPM PCB with the reflective foil and optically isolating black electrical tape removed. Middle: a fully constructed SIPM + scintillator assembly. Right: The SiPM PCB connected to the main PCB resting on the rails of the aluminum enclosure.}
  \label{fig:construction}
 \end{center}
\end{figure}

\subsection{SiPM PCB and scintillator}
The light collection system is designed around a 6$\times$6~mm$^2$ SensL C-Series MicroFB 60035 SiPM~\cite{sensl} (shown as element [h] in Fig.~\ref{fig:pcb}) due to its low operating voltage, compactness, single-photon sensitivity, temperature stability, and low cost. The \textit{standard output} of the SiPM creates a pulse that has a rise time of few nanoseconds and subsequently decays away over a few hundred nanoseconds.  The SiPM PCB is used to interface the SiPM to the scintillator and provides bias filtering to the SiPM.

In order to maximize the amount of light observed during each event, the SiPM is optically coupled to the scintillator with optical gel, wrapped in a reflective foil, and optically isolated using black electrical tape. In the left side of Fig.~\ref{fig:construction}, the reflective foil has been removed to show how the SiPM PCB is secured to the scintillator via four No. 0 screws. The SiPM and scintillator assembly connects to the main PCB via a 6-pin header and socket (element [m] and [e] of Fig.~\ref{fig:pcb}) and is secured in place using two aluminum hex standoffs (element [n]). The full assembly slides into place on the rails of the aluminum enclosure, as shown in the right side of Fig.~\ref{fig:construction}.

\subsection{Main PCB}
The primary purpose of the main PCB (the larger PCB shown in Fig.~\ref{fig:pcb}) is to perform the signal processing and data acquisition of the detector. The PCB consists of four circuits which are outlined on the silkscreen: the DC-DC booster (used to bias the SiPM), the non-inverting amplifier, the high-speed peak detector~\cite{peakdetector}, and the Arduino circuit. A detailed description of the circuit can be found in Section~\ref{sec:circuit}.

The Arduino Nano analog pins are used to perform the signal measurement (pin A0), control the OLED screen (pin A4 and A5), and measure the local temperature (the temperature sensor is show as element [d] in Fig.~\ref{fig:pcb} and connects to the A3 pin on the Arduino). The digital pins are used to read and write to/from the microSD card circuit (pins D10 - D13), monitor for a coincidence signal from a connected detector (pin D6), and pulse an LED (connected to pin D3) when a signal in the detector passes a software defined threshold. The USB Mini connection on the Arduino is used to update the Arduino code, power the detector, and record data to the computer.

The BNC receptacle (element [a]) on the top of the main PCB serves two purposes. First, it is connected directly to the SiPM anode and therefore allows users to use the raw SiPM pulse as a trigger for other data acquisition systems. This can be useful for using the scintillator and SiPM in conjunction with another system, such as a portable triggering system for a beamline measurement. An example of this is shown in Section~\ref{sec:measurements}. Secondly, we use the BNC connection to inject waveforms from an arbitrary waveform generator into the circuit to determine the response of the circuit. This is used to determine the relationship between the average measured ADC value and the SiPM pulse amplitude. 

The 3.5~mm female audio jack (element [c]) located on the back side of the main PCB is used to link multiple detectors together such that they can operate together in coincidence mode by setting them to either \textit{master} or \textit{slave} mode. The \textit{slave} is defined as the detector which records an event only if the \textit{master} detector also triggered within roughly 30~$\mu$s. Once two detectors are connected via a 3.5~mm male-to-male audio cable, the mode is selected by resetting both detectors using the reset button (element [b]). The first detector to be reset becomes the \textit{master}, while the second detector (if reset anytime between 10~ms and 2000~ms after the \textit{master}) becomes the \textit{slave}.  The tip conductor of the audio cable provides power to the other connected detector, which means that only one detector needs to be connected to a USB power source. This coincidence mode configuration is used for several of the example measurements shown in Section~\ref{sec:measurements}.

\subsection{SDcard PCB}
The microSD card reader/writer on the SDcard PCB (the smaller PCB shown in Fig.~\ref{fig:pcb}) is used as one of the methods to record data. This is particularly useful when the detector does not have access to a computer such as outdoors, on an airplane, in a hot air balloon, or on a rocket. In order to use the microSD card, dedicated Arduino code must be upload to the detector. When running the microSD card Arduino code, each time the detector is reset a new file is created on the microSD card with the naming format ``FileXXXO.txt," where ``XXX" is a 3-digit number that counts sequentially upwards, and the ``O" indicates the mode (``M" for \textit{master}, ``S" for \textit{slave}). The code allows up to 200 files to be generated in either mode, after which the detector will delete the previous 10 files and start counting again at file number 190. The SDcard PCB is mounted on the bottom of the main PCB via an 8-pin connector, shown as elements [j] in Fig.~\ref{fig:pcb}).

\section{Circuit description}\label{sec:circuit}
This section describes the various components of the circuit in detail. A schematic of the circuitry for each PCB is outlined in solid blue in Fig.~\ref{fig:circuit}. A high resolution image can be found in the supplementary material (found in /Pictures). 

\begin{figure}[h]
\begin{center}
\includegraphics[width=1\columnwidth]{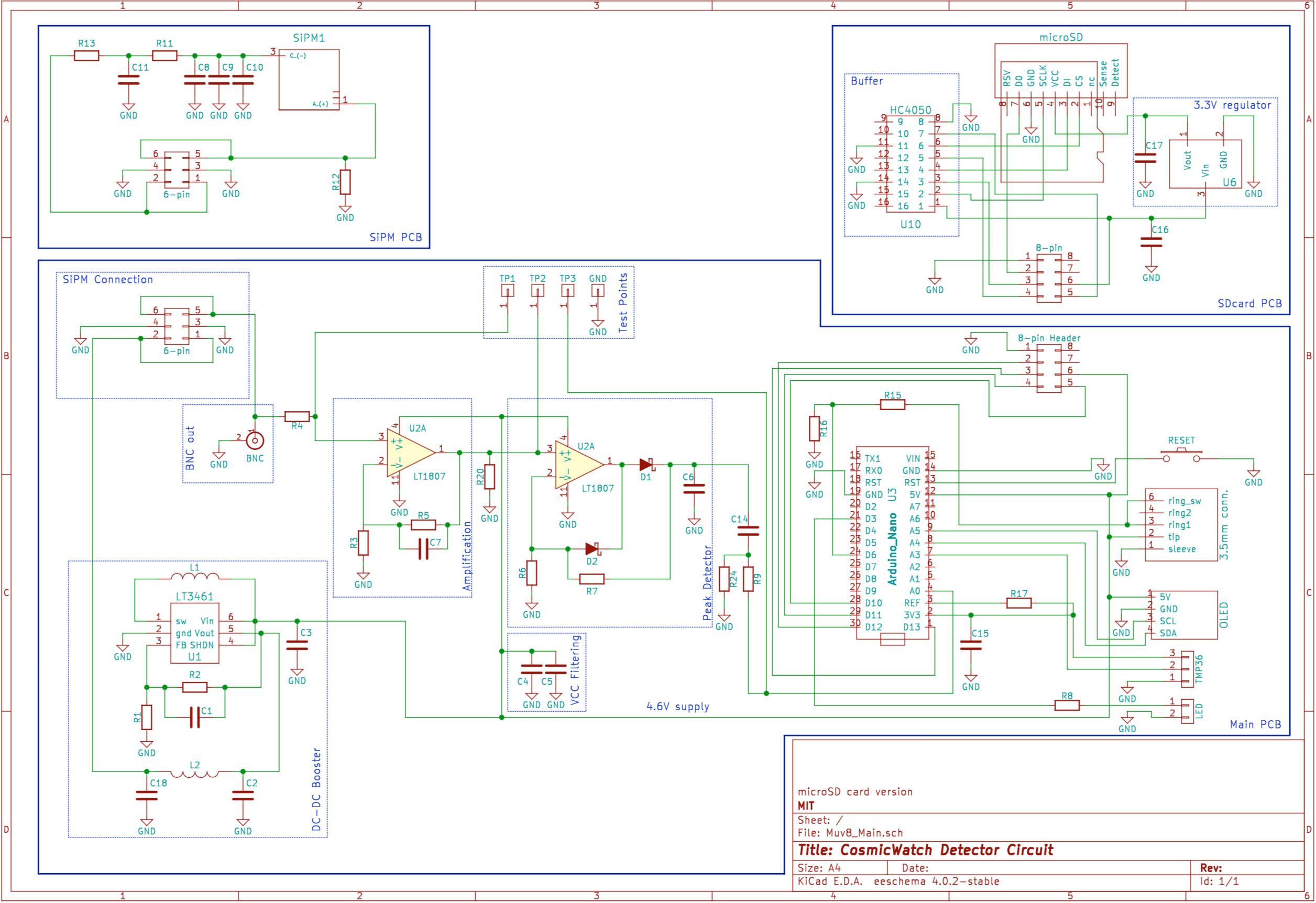}
\caption{The circuit diagram for the three PCBs. PCBs are outlined in solid blue, while various circuit blocks are outlined in dashed blue. The reference list for all components can be found in the supplementary material (SMT\_reference.xlsx). The circuit was designed using the open source program KiCAD~\cite{kicad}.}
  \label{fig:circuit}
 \end{center}
\end{figure}

The detector is powered through a USB Mini cable which plugs directly into the Arduino Nano. The USB provides a constant 5~V supply, which then passes through a built-in Schottky power rectifier which has a voltage drop of approximately 0.38~V~\cite{schottky}. The voltage after the rectifier is therefore approximately 4.6~V and is used as the Arduino analog reference as well as the supply voltage for the rest of the circuit. This voltage is filtered with three bypass capacitors (C3, C4, C5).

The DC-DC booster, shown on the bottom left of Fig.~\ref{fig:circuit}, is used to provide the biasing voltage for the SiPM. Here, the 4.6~V supply is boosted to +29.5~V and sent through a 6-pin connector to the SiPM PCB, which is used to overcome the breakdown voltage of the SiPM (+24.7~V) and apply an overvoltage of approximately 4.8~V. The resistors and capacitors connected to the cathode side of the SiPM (R13, R11, and C8, C9,C10, C11 in the top left of Fig.~\ref{fig:circuit}) are used as a low-pass filter to reduce noise from the DC-DC booster. The resistor, R12, is connected to the anode of the SiPM and holds the line at ground when there is no signal. 

When a micro-cell in the SiPM discharges, the generated current is sent through two of the pins on the the 6-pin connector to the main PCB. The signal branches into three paths: one path leading to the BNC receptacle, another to the test point 1 connection (element [f] in Fig.~\ref{fig:pcb} and labelled as TP1 in Fig.~\ref{fig:circuit}), and the last path to the input of the amplification circuit.  The non-inverting amplifying circuit uses the first stage of the  dual rail-to-rail LT1807 325~MHz op-amp~\cite{LT} (U2A). A capacitor (C7) is inserted in parallel with the feedback resistor (R6) to limit the frequency response of the amplification circuit and effectively obtains an amplification factor of approximately 25. The amplified waveform can be seen using test point 2 (TP2).

The amplified pulse enters the high-speed peak detector circuit (shown in the center of Fig.~\ref{fig:circuit}) which uses the second stage of the op-amp, U2A. When an amplified pulse enters this circuit, the output is driven to charge the capacitor C6 to the peak voltage of the input. When the pulse subsides, the Schottky diode D1 becomes back-biased and the capacitor discharges through resistors R6, R7, and R24. The time constant associated with the discharge of C6 through the resistors is approximately 0.3~ms. While the capacitor is discharging, the Arduino Nano samples the waveform several times and takes the average measurement as the measured ADC value. This measurement is then converted to an original SiPM pulse amplitude using the circuit calibration data described in the supplementary material (the Calibration section of ``Instructions.pdf").

The right side of Fig.~\ref{fig:circuit} shows the auxiliary components to the Arduino, including the schematics for the reset button, OLED screen, analog temperature sensor, and LED.

\section{Experimental Measurements}\label{sec:measurements}
This section describes various measurements that have been performed using the Desktop Muon Detectors. 

\subsection{Rate measurement 1km underground at Super Kamiokande  \label{sec:sk}}

\begin{figure}[!htb]
    \centering
    \begin{minipage}{.5\textwidth}
        \centering
        \includegraphics[width=1.\linewidth]{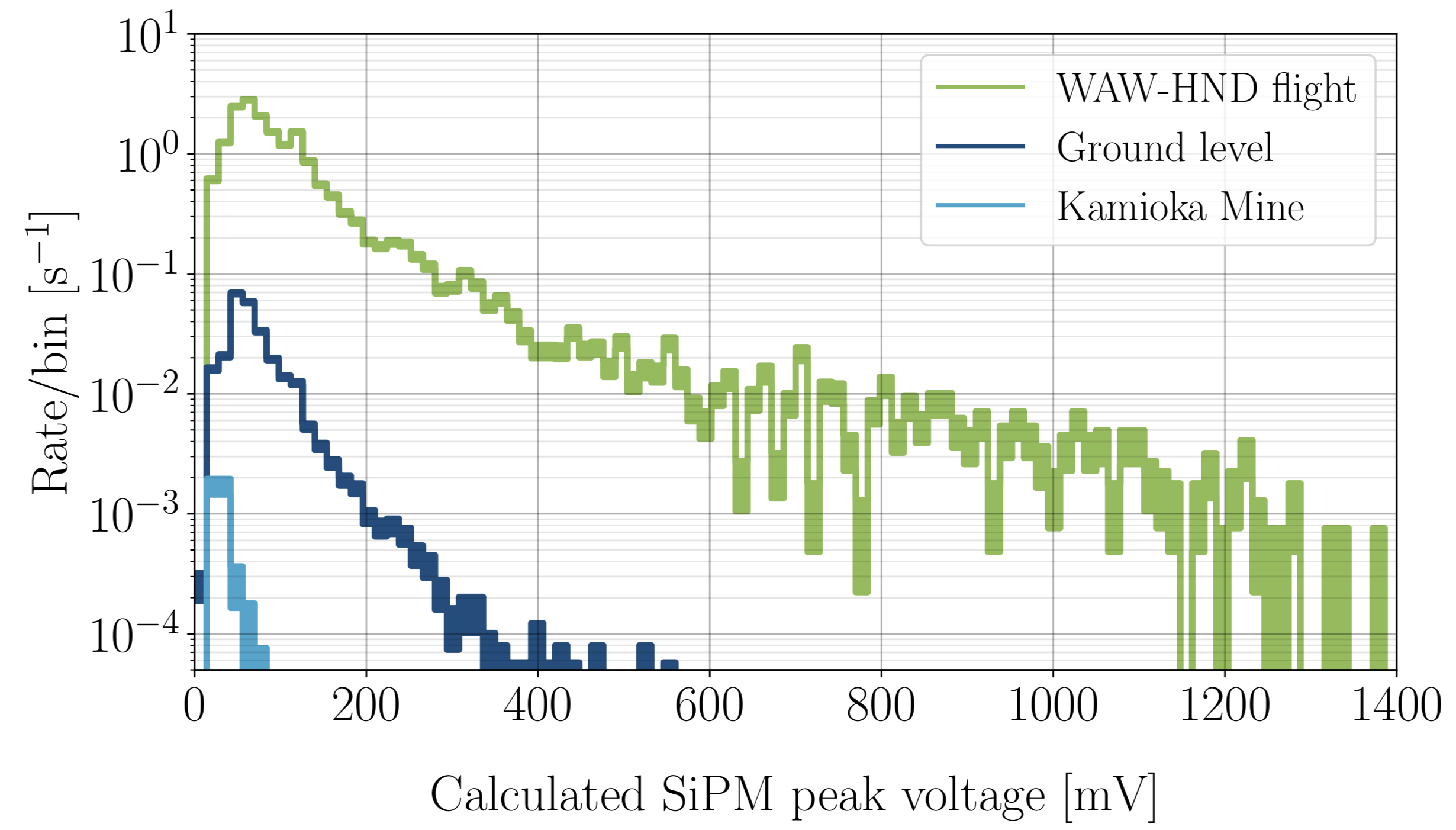}
    \end{minipage}%
    \begin{minipage}{0.5\textwidth}
        \centering
        \includegraphics[width=1.\linewidth]{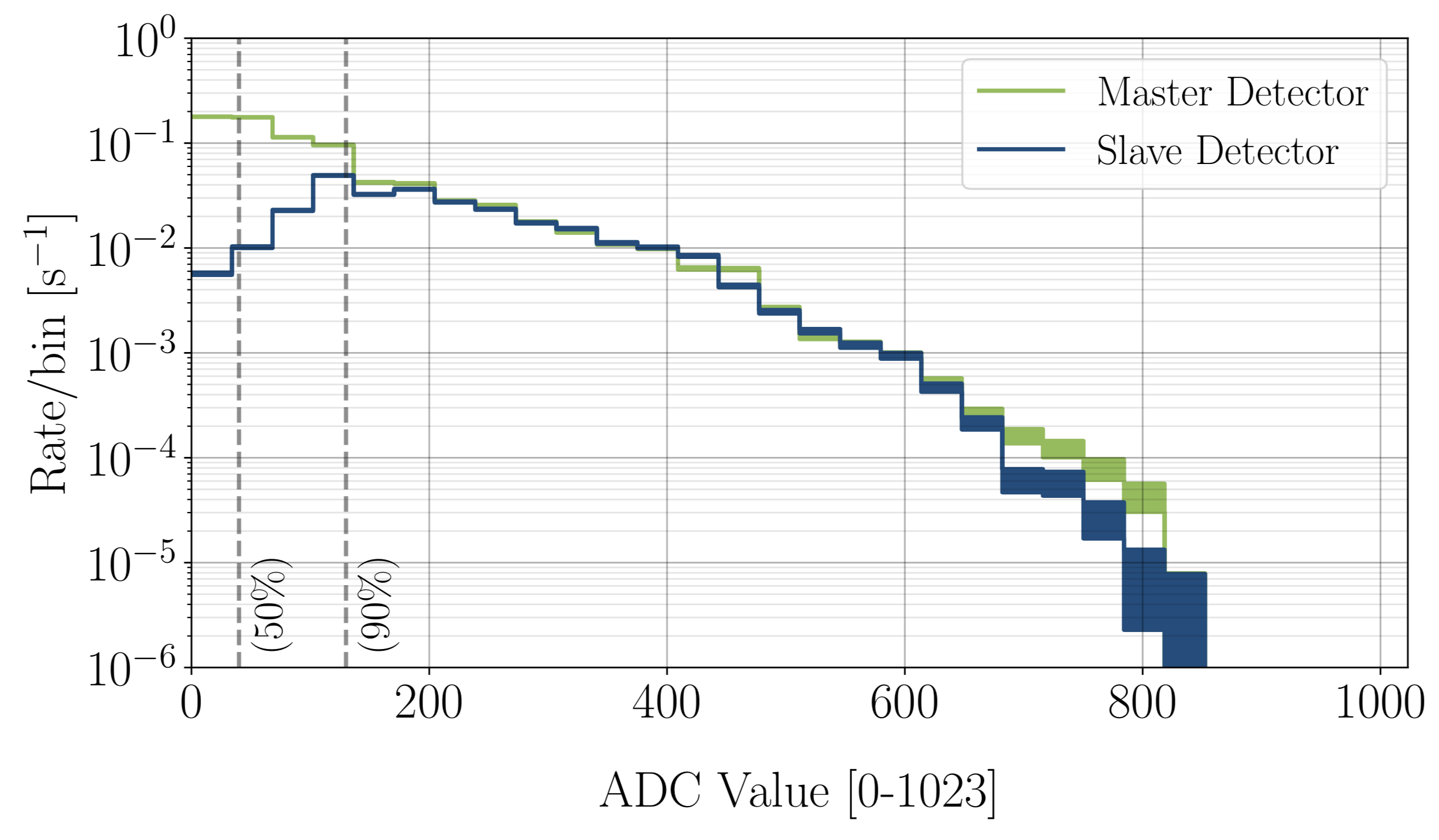}
    \end{minipage}
       \caption{(Left) The trigger rate as a function of calculated SiPM voltage for three locations:  in an airplane at 36,000ft, underground in the Kamioka Mine, and near sea level outside the mine. (Right) The measured ADC value for two detectors, removed from their aluminum enclosure and placed one-on-top of another (such that the scintillator is as close as possible) while connected in coincidence mode. The vertical dashed line shows the minimum ADC limit at which the trigger rate of the \textit{slave} is 50\% and 90\% that of the \textit{master}.}
       \label{fig:sk}
\end{figure}

\hspace{4ex} Two detectors were brought to the Kamioka Mine in Japan to perform a rate measurement near the Super-Kamiokande detector~\cite{sk}. Super-Kamiokande was built in the Kamioka mine and has roughly 2100~m.w.e. (meter-water equivalent) of overburden.  For this measurement, the detectors were left in their aluminum enclosure and placed one-on-top of each other. They were connected together using a 6-inch 3.5~mm audio cable and the \textit{slave} data was recorded directly to a laptop. Using the same detectors and setup, a rate measurement was also performed outside the Kamioka mine and in the airplane at 36,000~ft when travelling from Warsaw to Tokyo. Fig.~\ref{fig:sk} (left) shows the trigger rate of the \textit{slave} detector for these three measurements, as a function of calculated SiPM voltage. 

Fig.~\ref{fig:sk} (right) shows the measured ADC value for two detectors, removed from their aluminum enclosure and placed one-on-top of another such that the scintillators were in contact. The detectors were place in coincidence mode using a 6-inch 3.5~mm male-to-male cable. This measurement illustrates the ability of the detectors to distinguish between radiogenic backgrounds and cosmic ray muons. This data was recorded directly to a computer on the fourth floor lab of a ten floor office building. This measurement represents 48 hours of data. The vertical dashed line shows the minimum ADC limit at which the count rate of the \textit{slave} is 50\% (and 90\%) that of the \textit{master}. The \textit{slave} trigger rate was measured to be 0.281 +/- 0.001~s$^{-1}$.

\subsection{Portable trigger system for a beamline measurement at Fermilab \label{sec:energy}}
\begin{figure}[h]
\begin{center}
\includegraphics[width=0.6\columnwidth]{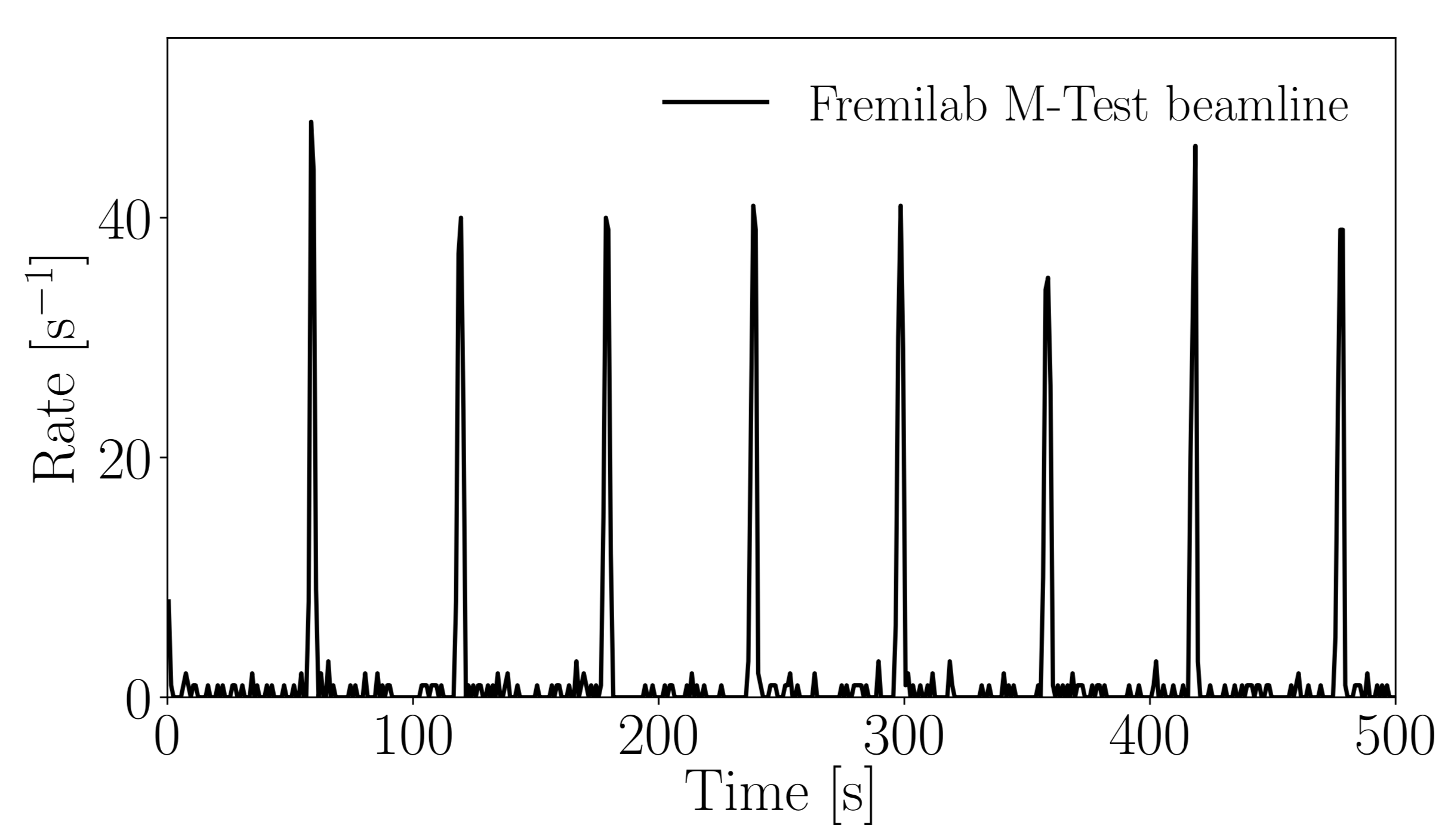}
\caption{The trigger rate as a function of time of a single detector placed in the Fermilab M-Test beamline. Here, the detector is triggering primarily on GeV-scale pions and electrons from the Fermilab Main Injector. }
  \label{fig:beam}
 \end{center}
\end{figure}

A single detector, powered by a 10000~mAh USB power bank, was placed in the Fermilab M-Test facility to trigger on secondary particles (GeV-scale pions and electrons) from the Main Injector. A BNC cable was connected from the detector to a triggering system for an IceCube experiment being run downstream. Fig.~\ref{fig:beam} shows the trigger rate as a function of time; the beam spills occur every minute for two seconds.

\subsection{Muon rate measurement at 33,000ft \label{sec:alt}}
A rate measurement was performed during a flight from the Boston International Logan Airport (BOS) to the Chicago O'Hare Airport (ORD) using a single detector. The data was recorded to the microSD cards with a single detector plugged into a 10,000~mAh USB power bank. The altitude of the airplane was collected from the flight records found in Ref.~\cite{flight_data}. 

\begin{figure}[h]
\begin{center}
\includegraphics[width=1\columnwidth]{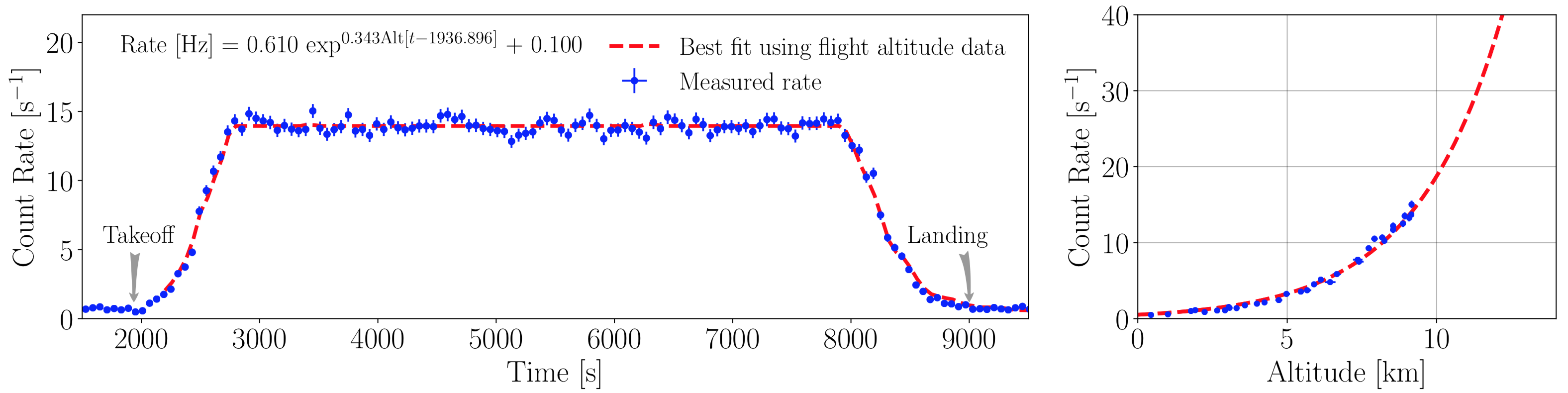}
\caption{(Left) The count rate measured during a flight from BOS to ORD as a function of altitude. The red-dashed line shows the actual amplitude of the airplane~\cite{flight_data} scaled by a fitted exponential shown at the top left of the plot. (Right) The count rate as a function of altitude.}
  \label{fig:flight}
 \end{center}
\end{figure}

Fig.~\ref{fig:flight}  (left) shows the trigger rate as a function of time. The altitude data was scaled by an exponential and fit to the count rate data. The best-fit equation is shown at the top of this figure. Fig.~\ref{fig:flight}  (right) shows the shows the altitude as a function of measured trigger rate. Here, we show the exponential fit extended beyond the measured values. The count rate uncertainties were calculated by taking the square root of the sum all the events measured at a particular altitude.

\subsection{Cosmic ray muon angular distribution measurement \label{sec:angle}}
\hspace{4ex} The final measurement illustrates the cosmic ray angular muon dependence. For this measurement, two detectors were set to coincidence mode and placed back-to-back, 52~mm apart, inside their aluminum enclosure. Placing the detectors back-to-back rather one-on-top of the other reduces the angular uncertainty. The angle was determined by securing the detectors in place on a 100~cm long rectangular bar and then positioning the bar against a wall at a known angle. We also positioned the detectors such that incoming trajectory of the muons had a minimum amount of attenuation from the building. Each data point represents approximately 10 hours of data. The measurement at $\theta = \pi/2$ is divided by a factor of 2, since at this angle it accepts cosmic-ray muons from both directions, whereas all the other angles only accept down going muons. 

\begin{figure}[h]
\begin{center}
\includegraphics[width=0.6\columnwidth]{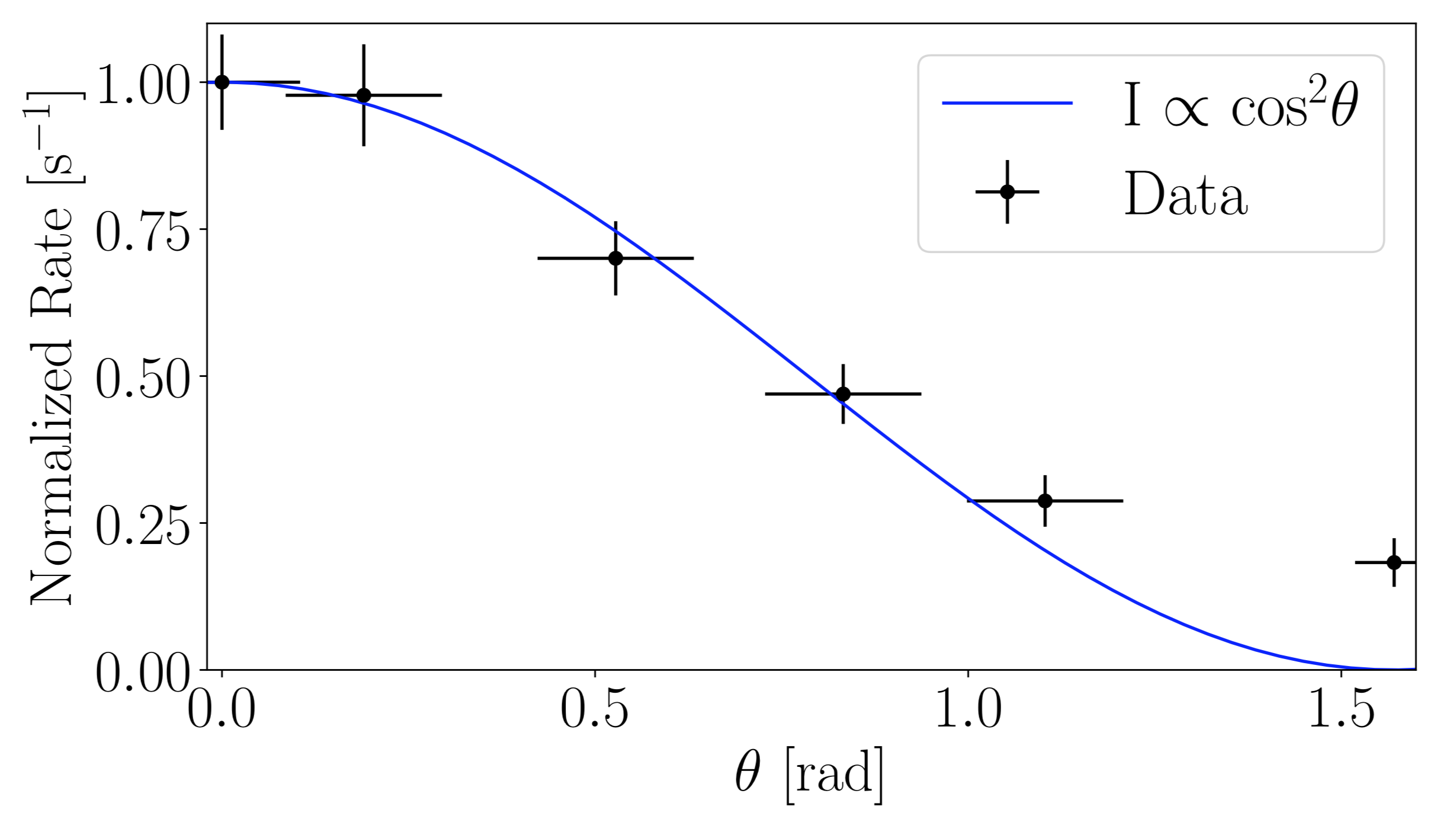}
\caption{The measured cosmic ray muon angular distribution measured by the \textit{slave} detector when connected in coincidence mode. The  prediction by the PDG is shown in solid blue.}
  \label{fig:angle}
 \end{center}
\end{figure}

The Particle Data Group (PDG) indicates that the angular dependence at sea level should follow a cosine squared dependence~\cite{PDG}. Fig.~\ref{fig:angle} shows the measured rate as a function of angle and a cosine squared distribution.

The angular uncertainty was calculated by looking the the maximum and minimum angle of a trajectory that could trigger both detectors. The angular error bars do not represent the uncertainty in the measurement, rather the uncertainty in the individual trajectory of the triggering muon. The uncertainty in the count rate was taken as the square root of the number of counts at a given angle.

\section{Conclusion}
\hspace{4ex} The CosmicWatch Desktop Muon Detectors are fully self-contained, low power, particle detectors that utilize modern technologies while remaining accessible to students. Participation in the project introduces students to a wide range of topics concerning, among other things, building and designing detectors, measuring cosmic rays muons, measuring radioactive sources, programming micro-controllers, developing data analysis techniques, learning about electronics, soldering, circuit design, creating software, data acquisition, and visualization techniques -- all at various levels of difficulty. 

The Desktop Muon Detector employs custom circuit boards that were designed to readout information from a silicon photomultiplier using an Arduino Nano. The detector includes a build-in microSD card reader/writer, a method to perform coincidence measurements, and software to record event information directly to computer or plot the data in real-time on the CosmicWatch website. Several measurements were performed using the detectors to illustrate their abilities under a variety of conditions; such as in a beamline, at high-altitudes, and underground with large overburdens. Supplementary material pertaining to building and testing a detector, as well as all the software required for operating the detector is provided in the  GitHub repository.

\acknowledgments
\hspace{4ex} This work was supported by grant NSF-PHY-1505858 (PI: Prof. Janet Conrad), MIT seed funding, and funds from the National Science Centre, Poland (2015/17/N/ST2/04064). The authors would like to thank SensL and Fermilab, for donations that made the development of this project possible, as well as P. Fisher at MIT and IceCube collaborators at WIPAC, for their support in developing this as a high school and undergraduate project. 

\bibliography{Master}
\bibliographystyle{ieeetr}
\end{document}